\documentclass[aps,pra,reprint,groupedaddress]{revtex4-2}

\usepackage{physics}
\usepackage{graphicx}
\usepackage{adjustbox}
\usepackage{ragged2e}
\usepackage[table]{xcolor}
\usepackage{hyperref}
\hypersetup{colorlinks,allcolors=black}

\begin{document}

% Deadline: June 15th we submit to full review of all authors. Mael will send an email after flipping noise figure with the new one

% Mael: Figur 7 redo with new Niccolo's suggested formating!

\title{High-Fidelity Quantum Entanglement Distribution in Metropolitan Fiber Networks with Co-propagating Classical Traffic}

\author{Matheus Sena$^{1}$}
\author{Mael Flament$^{2,3}$}
\author{Shane Andrewski$^{3}$}
\author{Ioannis Caltzidis$^{4}$}
\author{Niccolò Bigagli$^{3}$}
\author{Thomas Rieser$^{1}$}
\author{Gabriel Bello Portmann$^{3}$}
\author{Rourke Sekelsky$^{3}$}
\author{Ralf-Peter Braun$^{5}$}
\author{Alexander N. Craddock$^{3}$}
\author{Maximilian Schulz$^{1}$}
\author{Klaus D. Jöns$^{4}$}
\author{Michaela Ritter$^{1}$}
\author{Marc Geitz$^{1}$}
\author{Oliver Holschke$^{1}$}
\author{Mehdi Namazi$^{3}$}
\email{Corresponding author. mehdi@quconn.com}

\affiliation{$^{1}$Deutsche Telekom AG, Winterfeldtstraße 21, 10781 Berlin, Germany}
\affiliation{$^{2}$Qunnect NL BV, Elektronicaweg 10, 2628 XG Delft, Netherlands}
\affiliation{$^{3}$Qunnect Inc, 141 Flushing Ave, Suite 1110, Brooklyn, NY 11205, USA}
\affiliation{$^{4}$Institute for Photonic Quantum Systems (PhoQS), Center for Optoelectronics and Photonics Paderborn (CeOPP) and Department of Physics, Paderborn University, Warburgerstraße 100, 33098 Paderborn, Germany}
\affiliation{$^{5}$Orbit GmbH, Mildret-Scheel-Straße 1, 53175 Bonn, Germany}

\date{\today}

\begin{abstract}
The Quantum Internet, a network of quantum-enabled infrastructure, represents the next frontier in telecommunications, promising capabilities that cannot be attained by classical counterparts. A crucial step in realizing such large-scale quantum networks is the integration of entanglement distribution within existing telecommunication infrastructure. Here, we demonstrate a real-world scalable quantum networking testbed deployed within Deutsche Telekom's metropolitan fibers in Berlin. Using commercially available quantum devices and standard add-drop multiplexing hardware, we distributed polarization-entangled photon pairs over dynamically selectable fiber paths ranging from 10~m to 60 km, and showed entanglement distribution over up to approximately 100~km. Quantum signals, transmitted at 1324~nm (O-band), coexist with conventional bidirectional C-band traffic without dedicated fibers or infrastructure changes. Active stabilization of the polarization enables robust long-term performance, achieving entanglement Bell-state fidelity bounds between 85--99\% and Clauser-Horne-Shimony-Holt parameter $S$-values between 2.36-2.74 during continuous multiday operation. By achieving a high-fidelity entanglement distribution with less than 1.5\% downtime, we confirm the feasibility of hybrid quantum-classical networks under real-world conditions at the metropolitan scale. These results establish deployment benchmarks and provide a practical roadmap for telecom operators to integrate quantum capabilities.
\end{abstract}

\maketitle

\section{Introduction}
The field of quantum technology promises revolutionary applications ranging from quantum computing \cite{Cacciapuoti2020, Pogorelov2021, QuEra, Cacciapuoti2025} to enhanced metrology \cite{Komar2014, Wu2019, Kimble2008, Guo2020} and secure communications \cite{Bhaskar2020, Duan2001,Jasminder2021, Knaut2024, Kucera2024, Tagliavacche2025}, which will impact diverse areas such as financial transactions \cite{Aaronson2009a} and voting systems \cite{Yuan2008}. At the core of many of these applications lies entanglement, a distinctive quantum correlation that can be spatially separated, providing the foundation for quantum networks often referred to as the \textit{Quantum Internet}. Entanglement-based networks underpin key future capabilities such as quantum repeaters \cite{Briegel1998}, distributed quantum sensing \cite{Guo2020}, quantum computing \cite{Graham2022}, and time synchronization \cite{Quan2016}.

Photons, ideal quantum information carriers because of weak environmental coupling, often require or benefit from distribution through existing fiber optics to enable quantum applications. However, transmitting entangled states over a fiber infrastructure that already transmits classical data is a challenge. Key technical hurdles include the development of reliable and field-deployable quantum hardware, compatibility with legacy systems, and mitigation of noise, environmental fluctuations, and channel crosstalk. Previous demonstrations of quantum-classical coexistence have mostly been limited to single-channel \cite{strobel2024high,Thomas2023designing,mao2018integrating} or controlled laboratory settings \cite{thomas2024quantum,marcikic2004distribution,shen2022distributing}, with minor emphasis on long-term stability and multi-user metropolitan networks. Recently, more realistic approaches are starting to be explored \cite{rahmouni2024100, Yang2025, Kucera2024}. 

\begin{figure*}
\centering
\includegraphics[width =\textwidth]{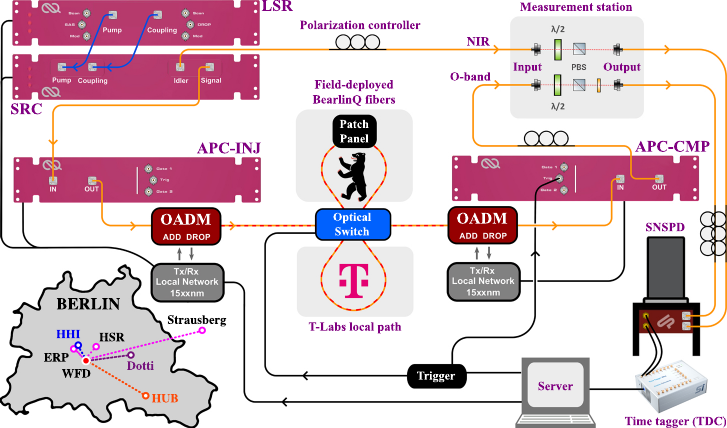}
\caption{\label{Fig:Experiment} \textbf{The BearlinQ layout}. Diagram of the setup for adaptive routing of polarization-entangled photon pairs between local (T-labs) and deployed fiber paths, shown in the lower left with a schematic of DT's Berlin fiber network nodes from the Winterfeldtstrasse laboratories (WFD). Polarization-entangled photons are generated at the SRC device optically driven by the LSR. The idler photon (795~nm) is routed to the measurement station, while the signal photon (1324~nm) travels through the APC-INJ/CMP stack, fiber-gapped by the intervening path-length segments, with the path selected by an optical switch. OADMs after the APC-INJ and before the APC-CMP allow co/counter-propagation of C-band channels. After the APC-CMP, signal photons go to the measurement station. Coincidences at given polarizations are logged using a Swabian Instruments TDC, while CHSH measurements are enabled by rotating $\lambda/2$ plates and polarizing beam splitters (PBS).}
\end{figure*}

The photonic degree of freedom used to encode entanglement is a key network design choice. The most popular are time-bin \cite{Tchebotareva2019}, frequency-bin \cite{Tagliavacche2025}, and polarization \cite{source2024, craddock2024automated}. Among these, polarization encoding is particularly appealing because of its ease of manipulation and its natural compatibility with other quantum systems. However, photon polarization is highly sensitive to birefringence fluctuations, which degrade entanglement fidelity and protocol reliability, an issue magnified in multipath topologies common to real-world networks. This makes active polarization compensation essential for stable entanglement distribution.

Here, we demonstrate the integration of high-fidelity polarization entanglement into an existing classical network under the city of Berlin. We call this hybrid quantum-classical network \textit{BearlinQ}, a name reflecting the technical robustness of our prototype while evoking the strength and resilience symbolized by Berlin's iconic bear. The network installation consists of distributing polarization-entangled O-band (1324~nm) photons over Deutsche Telekom's (DT) metropolitan fiber network in the presence of co- and counter-propagating C-band (1550~nm) classical traffic under real-world conditions. Robust quantum performance is enabled by Qunnect's integrated hardware suite comprised of an entanglement source, pumping lasers, and automatic polarization compensation to dynamically stabilize polarization across deployed fiber links. Integration with existing network infrastructure is made possible by optical add-drop multiplexers (OADMs), dynamic path switching, and adherence to International Telecommunication Union - Telecommunication Standardization Sector (ITU-T) channel standards, all without the need for dedicated fibers. To demonstrate network flexibility, we performed drift and switching experiments on different fiber paths. Long-term stability is confirmed by running the experiment over multiple days by alternating between local and deployed fiber segments of up to $\sim$60~km. Clauser-Horne-Shimony-Holt (CHSH) inequality \cite{Clauser1969} and fidelity measurements are periodically performed, confirming strong, continued quantum correlations in all scenarios. With automatic compensation, we maintain high fidelity with long-term quantum stability, demonstrating that operators can achieve reliable and scalable quantum networking on existing telecom infrastructure.

The paper is organized into two main sections that outline the network implementation and performance. Section \ref{sec:setup} describes the BearlinQ fiber testbed, covering key hardware elements such as the entanglement source, polarization stabilization, and quantum-classical signal integration. Section \ref{sec:exp} presents experimental results, including stability tests over urban distances, path-switching effects, and key metrics for network operators.

\section{Setup for Urban Entanglement Distribution}\label{sec:setup}
\subsection{The BearlinQ network}
Our quantum-classical hybrid network is centered around an entanglement generation hub connected to deployed fibers. A schematic of the layout is shown in Figure \ref{Fig:Experiment}. Entanglement is generated by an off-the-shelf commercial device (SRC, see Section \ref{Sec:FWMSource}) producing entangled photon pairs at 795~nm (NIR, \textit{idler}) and 1324~nm (telecom O-band, \textit{signal}), driven by a dedicated laser system (LSR). These rack-mounted (2U) devices enable deployment and comply with telecom collocation facility requirements. The 795~nm photons are routed through local fibers directly to a measurement station. The 1324~nm photons are directed to the same measurement station either along a short local fiber or through long-haul deployed fibers that exit the laboratory via a series of patch panels, traverse the BearlinQ network, and loop back to the lab. A MEMs fiber optical cross-switch allows dynamic selection of fiber paths. Classical communication signals required for coordination are transmitted over the C-band and combined with/separated from the quantum channel using OADMs. A central control server orchestrates the experiment and analysis.

To compensate for polarization drifts in the fibers caused by environmental fluctuations or switching events, the polarization state of the O-band photons is actively maintained by an automatic polarization compensator (APC). The APC consists in a pair of devices which, following entanglement generation, accept the quantum signal and interweave classical probing light at the same wavelength into the network fibers. This time-multiplexed classical signal enables multi-state polarization tracking and the application of dynamic corrections to mitigate polarization drifts and thus ensure that any arbitrary polarization will be well preserved during long-haul propagation (see Section \ref{Sec:APCDetails}). 

At the measurement station, O-band and NIR photons are analyzed. Detection is performed using superconducting nanowire single-photon detectors (SNSPDs). Arrival times are time-stamped with a time-to-digital converter (TDC) and processed at the server to extract the CHSH $S$-value. The measurement station, consisting of motorized half-wave ($\lambda/2$) plates, fixed polarizing beam splitters, and a bandpass filter to suppress C-band noise, supports automated CHSH tests. Manual controllers calibrate the local polarization reference and also ensure alignment with the polarization sensitive SNSPDs. Run-to-run variations in the measured $S$-values are primarily attributed to uncertainties in this manual calibration procedure (see Supp. Mat. \ref{Sec:SupMattPerf}).

\begin{figure}[h!]
\centering
\includegraphics[width=0.99\columnwidth]{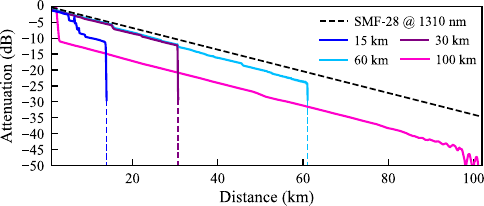}
\caption{\label{Fig:OTDR} \textbf{Fiber characterization.} OTDR measurements of fibers used individually or stitched together to reach $\sim15$~km (HHI), $\sim30$~km (Dotti), $\sim60$~km (2$\times$ Dotti), or $\sim100$~km (Strausberg) (specifically: 13.6, 30.5, 61.0 and 97.9~km). Trace discontinuities correspond to splices, connectors, and mating sleeves. For reference, the expected attenuation of standard SMF-28 single-mode fiber at 1310~nm is $\sim0.31$~dB/km, as indicated by the dashed black line.}
\end{figure}

Finally, we characterize the network by performing optical time-domain reflectometer (OTDR) measurements on each length of fiber. Explicit characterization is essential because real-world fibers are subject to a range of imperfections including environmental influences, splices, connector interfaces, and inconsistencies in manufacturing or installation that can introduce losses exceeding nominal specifications. This is evident in BearlinQ, where the 30~km and 60~km paths exhibit losses close to those expected, but the 15~km and 100~km stretches contain segments with greater attenuation (see Figure~\ref{Fig:OTDR}).

\subsection{Quantum-Classical Integration and Entanglement Source}
\label{Sec:FWMSource}
A key consideration in the design of networks that support both quantum operations and conventional classical traffic is the selection of relative wavelengths for the two domains, along with the closely related choice of entanglement source. This decision is primarily informed by two factors: spectral channel density and noise management. Assigning classical signals to the C-band and quantum signals to the O-band effectively addresses both concerns. The C-band’s high channel density enables support for a large number of simultaneous users through a mature and standardized dense wavelength division multiplexing (DWDM) infrastructure, featuring narrow 50GHz channel spacing. In contrast, the O-band is limited to coarse WDM (CWDM) standards, typically offering only three usable channels spaced by 20~nm, which significantly constrains its multiplexing capacity. From a telecom operator’s perspective, this disparity naturally favors allocating classical signals to the C-band, which is better suited to high user density, bandwidth-intensive applications, and diverse service requirements. Conversely, the O-band is better suited for low-noise quantum channels, which are more appropriate for transmitting fragile entangled states.

Single-photon signals are extremely susceptible to noise in the form of uncorrelated photons that make their way into the measurement devices. Once polarization is stabilized through the network, the relative rate of the quantum signal and of the noise background is the primary limiting factor to the quality of the measured entanglement as the distances increase. The dominant inter-band noise source in optical fibers is spontaneous Raman scattering~\cite{bloembergen1967stimulated}, which occurs when light interacts with its propagation medium generating broadband frequency-shifted photons that overlap with the quantum signal and thus cannot be filtered. Proper selection of the relative wavelengths of the quantum and classical channels can effectively mitigate this issue. Raman scattering is an asymmetric process, predominantly producing Stokes-shifted (longer wavelength) photons. Transmitting quantum in the O-band and classical data in the C-band is thus particularly favorable. By placing classical channels at longer wavelengths, the majority of Raman noise falls outside the quantum detection window, significantly avoiding in-band noise pollution. This approach enables the use of high classical launch powers, relaxes spectral filtering requirements (reducing insertion loss) and enhances the robustness of the system for deployment of coexisting channels over existing fiber infrastructure.

Having identified the O-band as the ideal window for quantum signals, we employ Qunnect's SRC as the source of entangled photons. It operates at room temperature delivering $>$$10^7$ photon pairs/s with $\sim$$1$~GHz linewidth (high brightness) and intrinsically stable polarization entanglement with large signal-idler cross-correlations ($g^{(2)}_{si} \gtrsim 30$) \cite{Davidson2021} and a heralding efficiency of $\sim 30-50\%$ \cite{source2024}. Its native O-band wavelength also removes the need for frequency conversion. These features make the SRC a suitable tool for the generation of entanglement in quantum networks run by telecom operators, as it eliminates many constraints of conventional sources such as limited brightness or fidelity, or the need for cryogenics, vacuum apparatus, or controlled laboratory environments. 

The device uses diamond-scheme spontaneous four-wave mixing (SFWM) in a warm rubidium $^{87}$Rb vapor cell to generate bichromatic polarization-entangled photon pairs \cite{source2024}. The LSR 780~nm \textit{pump} and 1367~nm \textit{coupling} lasers drive the atomic transition from the ground state, $|5S_{1/2}\rangle$, to the doubly excited state $|6S_{1/2}\rangle$. Decays to the $|5P_{1/2}\rangle$ and $|5S_{1/2}\rangle$ states emit photons at 1324~nm and 795~nm, respectively. The phase-matching condition ensures strong temporal and spatial correlations between photon pairs. By matching the polarization of the pump and coupling fields, the Rb spectroscopic structure guarantees the emission of photon pairs in the maximally entangled Bell state $\ket{\Phi_+} = \frac{1}{\sqrt{2}} (\ket{HH}+\ket{VV})$. 

The photon pair generation rate scales linearly with pump laser power, as the probability of pair creation is directly proportional to the number of pump photons available for the nonlinear interaction \cite{source2024}. Increasing the pump power is particularly useful in compensating for losses that accumulate with increasing fiber length, as a higher pair-generation rate helps maintain sufficient photon rates for use-cases. However, while increasing pump power helps overcome larger losses, it also introduces a trade-off in signal-idler correlation strength, as quantified by the cross-correlation function $g^{(2)}_{si}$ \cite{Davidson2021}. The inverse relationship between pair-rate and $g^{(2)}_{si}$ imposes a practical limitation, as large pump powers ultimately reduce the proportion of usable entangled pairs. To accommodate different fiber lengths while ensuring high-fidelity a constant coupling power of $8.4$~mW was used, along with variable pump powers of $\sim450-900$~\textmu W.

\subsection{Active Polarization Stabilization}
\label{Sec:APCDetails}
Environmental factors such as temperature fluctuations \cite{Ren1988} and mechanical stress \cite{Day1983} on single-mode, non-polarization maintaining optical fibers induce birefringence changes, causing unwanted rotations of the state of polarization of propagating photons. Protocols relying on polarization require precise control of these states to correctly analyze correlations. Uncompensated variations can effectively randomize the basis, degrading the fidelity and usability of entanglement across the network.

To ensure accurate correlation measurements, we actively track and compensate for polarization drifts by maintaining a stable and distributed polarization reference across all nodes. The APC system (injector, APC-INJ, and compensator, APC-CMP) provides such a reference and stabilizes the fibers to it. The APC-INJ transmits a known control sequence of approximately the same wavelength as the qubits, time-multiplexed with the quantum channel, to the APC-CMP at the opposite end of a fiber link (see Supp. Mat. \ref{Sec:SupMattPerf}). The APC-CMP analyzes the deviations of a sequence of states from their intended polarizations and applies corrections through a low-loss and rapid feedback mechanism \cite{craddock2024automated}.

\begin{figure}[h!]
\centering
\includegraphics[width=\columnwidth]{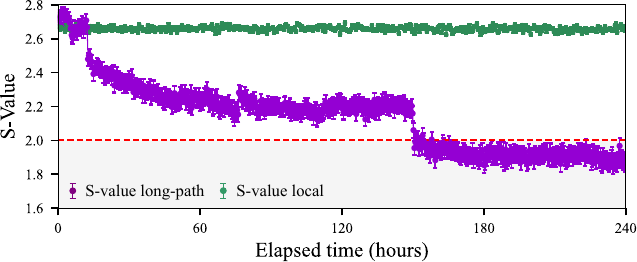}
\caption{\label{Fig:driftanalysis} \textbf{Passive operation}. Time evolution representative example of the CHSH $S$-value for both the local path (green) and a 30~km deployed fiber (purple), measured without active polarization stabilization \cite{Sena2024}. The common fiber segments flanking the APC pair, as well as the local short fiber, exhibit exemplary high stability. In contrast, the deployed path shows a gradual decline in $S$, with occasional abrupt drops, highlighting the dominant impact of environmental perturbations on long-term stability. The red line marks the threshold at $S = 2$: the classical bound for local hidden variable theories. $S > 2$ indicates entanglement, but does not guarantee high quantum channel fidelity. Error bars denote $1\sigma$ uncertainties derived from Poissonian propagation.}
\end{figure}

\begin{figure*}[t!]
\centering
\textcolor{orange}{\textbf{Idler WP=$0^\circ$}}
\textcolor{blue}{\textbf{Idler WP=$22.5^\circ$}}
\textcolor{red}{\textbf{Idler WP=$45^\circ$}}
\textcolor{cyan}{\textbf{Idler WP=$67.5^\circ$}}\\
\vspace{0.75em}
\begin{minipage}[t]{0.49\textwidth}
 \centering
 \includegraphics[width=\linewidth]{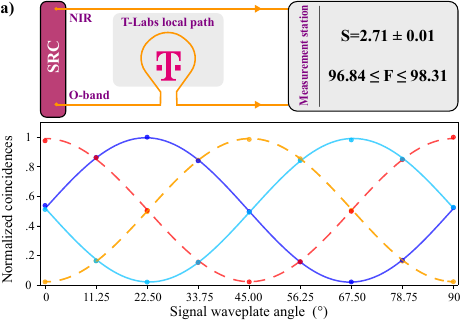}
\end{minipage}
\hfill
\begin{minipage}[t]{0.49\textwidth}
 \centering
 \includegraphics[width=\linewidth]{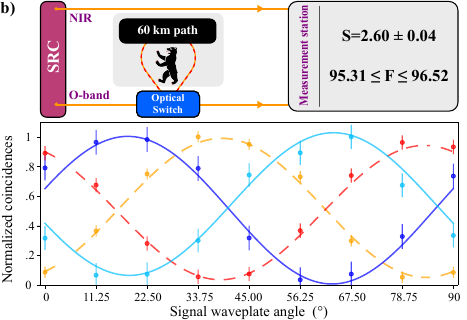}
\end{minipage}\\
\vspace{0.75em}
\begin{minipage}[t]{0.49\textwidth}
 \centering
 \includegraphics[width=\linewidth]{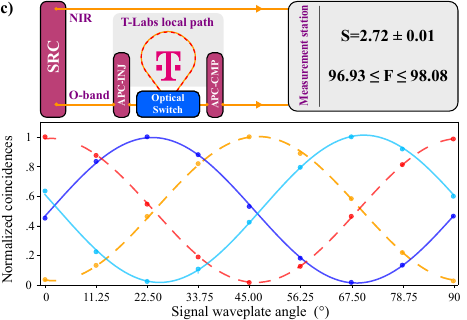}
\end{minipage}
\hfill
\begin{minipage}[t]{0.49\textwidth}
 \centering
 \includegraphics[width=\linewidth]{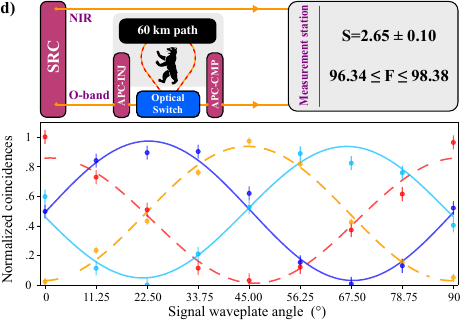}
\end{minipage}
\caption{\label{Fig:CHSH-Grid} \textbf{CHSH measurements across BearlinQ}. Normalized coincidence counts as a function of signal waveplate angle for idler waveplate settings of $0^{\circ}, 22.5^{\circ}, 45^{\circ}$, and $67.5^{\circ}$ in four configurations: \textbf{(a)} local fiber loop without the optical switch nor the APC; \textbf{(b)} 60 km fiber loop including a switch and patch panels without the APC; \textbf{(c)} local fiber loop including the APC, the switch, and the patch panel; and \textbf{(d)} 60 km fiber loop including a switch, patch panels and the APC. Error bars represent $1\sigma$ uncertainties from Poissonian count statistics propagated through the CHSH $S$ calculation. The integration time for each waveplate configuration was 1 s (2 s) for the local (60 km) path.}
\end{figure*}

We characterize the fibers to assess the drift rate and evaluate the required period of basis realignments by performing CHSH measurements and quantum fidelity, $F_Q$, over both local and deployed fibers in the absence of active stabilization. The $S$-value is calculated using standard methods \cite{semenov2010entanglement}, while the upper/lower quantum fidelity bound calculation is detailed in Supp. Mat. \ref{Sec:AppendixFidelity}. Figure~\ref{Fig:driftanalysis} illustrates the necessity of automatic polarization stabilization in field-deployed optical fiber networks and establishes the fundamental premise for the studies presented in this work. It shows the results of a long-term entanglement distribution experiment presented in \cite{Sena2024}, which used a setup similar to that proposed in this work, but without active stabilization. As observed, the temporal evolution of the $S$-value and $F_Q$ for the selected fiber exhibits a gradual drift over time, punctuated by sporadic sudden jumps. This behavior is consistently observed across different deployments and is likely attributable to common installation practices. In this particular case, all fibers are buried, and the infrastructure is modern and actively maintained by DT. Nonetheless, our findings confirm that active polarization stabilization is essential to ensure continuous and stable network operation for telecommunication providers.

\section{Entanglement Distribution Through Berlin}\label{sec:exp}
\subsection{CHSH over Local \& Urban Distances}
To characterize our network, we measure the CHSH inequality violation and the quantum state fidelity relative to the $\ket{\Phi_+}$ Bell state. We establish a baseline for the performance of the entanglement distribution both locally and over a 60~km fiber link without APC, nor path routing. The local CHSH test yields an $S$-value of \( S = 2.71 \pm 0.01 \) and a quantum fidelity of \( 96.84\% \leq F_{Q} \leq 98.31\% \) (Figure~\ref{Fig:CHSH-Grid} \textbf{a}). The same measurement was performed adding the switch and over 60~km resulting in a $S$-value of \(S = 2.60 \pm 0.04 \) and \( 95.31\% \leq F_{Q} \leq 96.52\% \) (Figure~\ref{Fig:CHSH-Grid} \textbf{b}). The fidelity reduction at 60~km stems from the increased pair generation rate required to overcome loss, which in turn degrades signal-idler correlations.

Flexible path routing is necessary for network operation in accordance with current classical networks and future use-cases. To enable routing, we introduce automated path-switching in our setup. The APC re-aligns polarization after each path change, ensuring high entanglement fidelity regardless of the route. To confirm that the additional hardware does not negatively affect the results, we repeat the measurement locally with all the devices in place. An $S$-value of \( S = 2.72 \pm 0.01 \) is measured, consistent with the initial local results (Figure ~\ref{Fig:CHSH-Grid} \textbf{c}). After switching to the 60~km path and running a compensation routine, we measure $S = 2.65 \pm 0.10 $ and quantum fidelity of \( 96.34\% \leq \text{F}_{Q} \leq 98.38\% \) (Figure ~\ref{Fig:CHSH-Grid} \textbf{d}), once again lower due to higher pair rates. These results demonstrate that automated path-switching and polarization compensation function as intended, eliminating the need for manual adjustments and enabling stable quantum correlations over independent fiber paths.

The consistency of CHSH curves depends on the calibration procedure which begins by inserting perpendicularly aligned polarizers in the SRC to preselect the polarization state $\ket{H}$ or $\ket{V}$. Motorized $\lambda/2$-waveplates are then aligned with their slow/fast axes set to the zero-angle position, which is then stored as a zero-position reference. Next, manual polarization controllers before the measurement station are adjusted to minimize single-photon counts, thus defining the reflected basis of the PBS's as the reference basis of measurement. Subsequently, the waveplates are rotated to the diagonal basis, where LCR voltages are then fine-tuned to minimize coincidences. These steps could be automated, eliminating user-induced variability and enabling more precise and repeatable calibrations than those presented herein.

\subsection{Sustained Automated Path Switching}\label{sec:Switching}
\begin{figure*}[t!]
\centering
\begin{minipage}[t]{0.451\textwidth}
\vspace{-28.9em}
\includegraphics[width=0.97\linewidth]{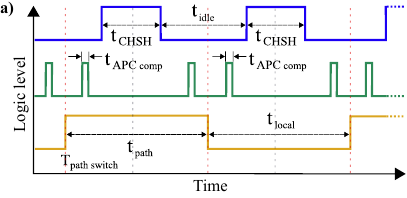}

\vspace{-1em}
\includegraphics[width=\linewidth]{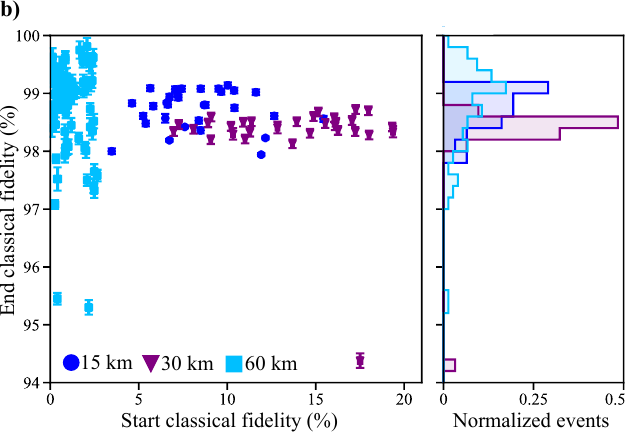}
\end{minipage}
\begin{minipage}[t]{0.542\textwidth}
 \centering
\includegraphics[width=\linewidth]{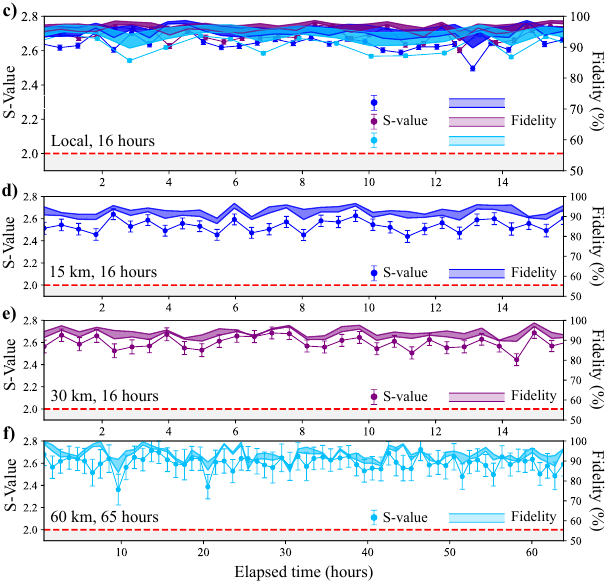}
\end{minipage}
\caption{\label{Fig:multipathswitch} \textbf{Sustained automated multi-path switching in a metropolitan network.} \textbf{a)} Sequence of logic operations showing key timing intervals for CHSH measurement (blue, \(t_{\mathrm{CHSH}}\)), active polarization control (green, \(t_{\mathrm{APC\,comp}}\)), path switching (yellow, \(T_{\mathrm{path\,switch}}\)), and local acquisition (yellow, \(t_{\mathrm{local}}\)). Vertical dotted lines mark switching events. \textbf{b)} Start and end classical fidelity for all APC routines during the switch from BearlinQ to local fiber. Histograms at right show the distribution of final classical fidelities measured by the APC. \textbf{c)} Local long-term CHSH and quantum fidelity traces, each corresponding to a switch from one of the deployed fibers (see following panels); similar color tones indicate the same measurement run. \textbf{d), e), f)} Long-term CHSH and quantum fidelity over 15~km, 30~km, and 60~km, respectively. Error bars in \textbf{c--f} show $1\sigma$ uncertainties from Poissonian count statistics propagated through the CHSH $S$ calculation. In \textbf{c--f}, the red dashed line marks $S=2$. The integration time for each waveplate configuration was 1 s for the local, 15 km, and 30 km paths, and 2 s for the 60 km path.}
\end{figure*}

A second benchmark to establish BearlinQ as a exemplary network is the demonstration of long-term reliable operation. This is shown by continuously operating the system for multiple days, including automated switching across the available paths when integration times allowed. To assess the robustness of the adaptive routing protocols, periodic CHSH and fidelity measurements are performed while alternating the photon route between the local T-lab path and the long-distance fibers. Each time the fiber path is switched, the APC compensation is triggered. A timing diagram of the routine is shown in Figure \ref{Fig:multipathswitch} \textbf{a}. With the addition of further switch ports, paths, and entanglement sources, this architecture could scale towards a multi-user quantum routing framework. 

The core of our reliability tests are long-term measurements for 15~km, 30~km, and 60~km paths over durations ranging from 16 to 65 hours. All measurements are performed in parallel with periodic switching between configurations and active compensation dynamically adapting to varying fiber conditions (Figure 5). 

To illustrate the effectiveness of polarization compensation, Figure~\ref{Fig:multipathswitch} \textbf{b} shows the classical APC-calculated \textit{start fidelity} (pre-compensation) and \textit{end fidelity} (post-compensation) for each switching event. We see that different fiber paths require distinct polarization transformations, resulting in variable compensation times and at the cost of network downtime, which ranges from tens of milliseconds to a few seconds (average compensation time of 511~ms across all paths of Figure~\ref{Fig:multipathswitch}). The $95^{\text{th}}$ percentile compensation times for a change from local to 60~km path and vice versa are $9.5$~s and $250$~ms, respectively. 

Figure ~\ref{Fig:multipathswitch} \textbf{c}, \textbf{d}, \textbf{e}, and \textbf{f} and show the calculated $S$-values and quantum fidelity bounds for each fiber length employed in the path-switching experiment. We observe sustained quantum correlations across all configurations, with an average $S = 2.66 \pm 0.04 $ and $90\% \leq F_Q \leq 99\%$ whenever measuring the local path. For the 15~km, 30~km, 60~km loops, we measure average CHSH parameters of $S = 2.53\pm0.05$, $S = 2.61\pm0.06$, and $S = 2.61\pm0.09$, respectively. On the 15~km path, the quantum fidelity lower bound exceeded 90\% during 75\% of the runtime, while on the 30~km and 60~km fibers this threshold was maintained for 95\% and 79\% of the runtime, respectively. All paths exceed the 85\% quantum fidelity lower bound 100\% of the time, except the 60 km path which does 98.5\% of the time. Collectively, these results demonstrate robust long-term operations and the flexibility of our network.

\begin{figure}[h!]
\centering
\includegraphics[width=\columnwidth]{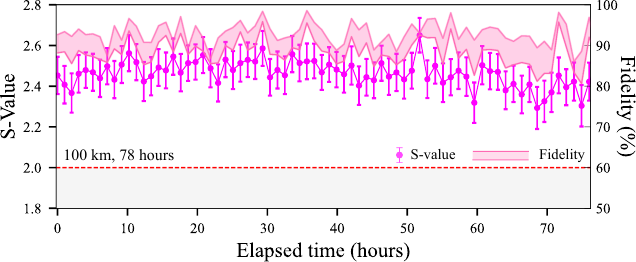}
\caption{\label{Fig:100km} Entanglement distribution over 100~km through Berlin. These results demonstrate the system’s ability to maintain strong quantum correlations despite significant transmission losses and persistent uncontrolled environmental perturbations in the deployed fiber. This dataset was acquired without the path switching equipment included. Each waveplate setting in the CHSH and fidelity measurements was integrated for 30~s.}
\end{figure}

To test the limits of our testbed, we also run fidelity and CHSH measurements over 100~km. However, due to the need for longer integration times, this data was acquired without path switching nor DWDMs. Figure \ref{Fig:100km} shows the $S$-values and quantum fidelity bounds for the 100~km loop run continuously for 78~hours. In this experiment, the system achieved $S=2.58 \pm 0.03$ and a fidelity between $80.5\% \leq F_Q \leq 98.4\%$.

\subsection{Operational Metrics}
Network operators rely on a set of key performance indicators (KPIs) to evaluate service performance. In conventional telecom networks, these include uptimes exceeding $99.99\%$, negligible interchannel interference, and consistent signal integrity. Although particularly challenging, translating these expectations into the quantum domain is essential for building practical hybrid quantum-classical networks. Achieving telecom-grade reliability in quantum systems requires active stabilization, real-time diagnostics, and ultra-low-noise operation, all nontrivial engineering feats. Our system addresses these challenges, taking a step toward transitioning quantum networks from experimental setups to real-world infrastructure. It demonstrates quasi-uninterrupted entanglement distribution over metropolitan-scale infrastructure under real operating conditions.

Operator-aligned metrics are achieved, including low downtime, negligible impact to classical traffic, high entanglement pair-rates, and high-fidelity transmission. Here, downtime is defined as periods when $F_Q$ drops below $85\%$ or when recalibration is needed. Automatic stabilization allows for near carrier-grade reliability, as shown by $<1.5\%$ downtime on a 60~km path while maintaining $F_Q > 85\%$ (Figure~\ref{Fig:multipathswitch}\textbf{d}). Fidelity was mainly limited by the high entanglement generation rate and detuning of the APC (see Supp. Mat.~\ref{Sec:AppendixFidelity}). A further minor reduction in fidelity of $0.3\%$ is attributed to polarization-dependent loss (PDL) of $\sim0.9$~dB (excluding fibers), from the optical switch and WDM components.

Practical quantum networking also demands useful pair-rates. Instrumental for this result was the high SRC pair generation rate of $\geq10^7$ pairs/s. Our deployed system consistently delivers coincidence rates ranging from $5\times10^2-10^7$ entangled pairs/s after losses, depending on the selected fiber path. However, in the absence of ultralow-loss fibers or quantum repeaters, the attenuation of the O-band fiber at $\sim$0.31~dB/km limits practical point-to-point deployments to distances of $\sim$100~km. Previous O-band demonstrations relied on fiber spools in laboratory settings \cite{marcikic2004distribution,shen2022distributing} or short campus loops \cite{rahmouni2024100}. Here, we exceed those benchmarks by reaching a rate of $\geq10^2$ pairs/s using up to 100~km of urban fiber ($-45$~dB attenuation), demonstrating a resilient platform for useful quantum networking in the real world. 

Table \ref{Tab:comparison} summarizes key performance metrics and highlights how this deployment establishes a new operational benchmark for metropolitan quantum networks.

\renewcommand{\arraystretch}{1.4}
\begin{table*}[t!]
\begin{adjustbox}{max width=\textwidth}
\begin{tabular}{|c|l|c|c|c|c|c|c|c|}
\hline
\rowcolor{gray!5}
\textbf{Reference} & \textbf{Location} & \textbf{Setting} & \textbf{Fiber (km)} & \textbf{Fidelity (\%)} & \textbf{Pair Rate (pairs/s)} & \textbf{Quantum $\lambda$ (nm)} & \textbf{Classical $\lambda$ (nm)} & \textbf{Duration} \\
\hline
\textit{This work} & Berlin, DE & Urban & $10^{-2}$ to 100 & 99-85 & 10$^7$- $5\times 10^2$ & 795, 1324 & 1560.61 & days (ea.) \\
\cite{thomas2024quantum} & Chicago, US & Laboratory & 30.2 & $>90$ & - & 1290-1310 & 1547 & - \\
\cite{marcikic2004distribution} & Geneva, CH & Laboratory & 50 & $\sim 80$ & $\sim10^1$ & 1310 & 1534 & hours \\
\cite{shen2022distributing} & Singapore, SG & Laboratory & 50 & 92.5-97 & $10^4$ & 586, 1310 & - & - \\
\cite{craddock2024automated} & Brooklyn, US & Urban & 34 & 99 & $5\times 10^5$ & 795, 1324 & - & 15 days \\
\cite{strobel2024high} & Stuttgart, DE & Urban & 35.8 & 94.5 & $\sim10^3$ & 780, 1515 & - & days \\
\cite{Thomas2023designing} &Chicago, US & Urban & 47.9 & $\sim 99$ & $\sim 10^2 $ & 1280-1318 & 1549-1565 & - \\
\cite{mao2018integrating} & Shandong, CN & Urban & 66 & N/A & N/A & 1310 & 1490, 1538 & - \\
\hline
\end{tabular}
\end{adjustbox}
\caption{\label{Tab:comparison} Comparison of performance metrics of O-band quantum entanglement distribution or comparable testbeds\footnote{\noindent \textit{To our knowledge these represent the state-of-the-art of O-band or bichromatic photon-pair entanglement distribution. We distinguish two types of experiments: fiber spool-based in-lab setups or nearby buildings supplemented with spools (\textit{laboratory}), and real-world deployments using installed fiber (\textit{urban}). If a metric was unavailable, we used ``-"; if not relevant, we used ``N/A".}}}
\end{table*}

\begin{figure}[h!]
\centering
\includegraphics[width=\columnwidth]{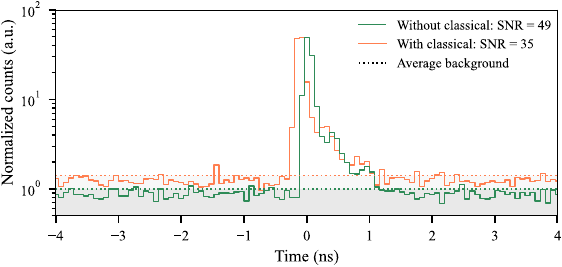}
\caption{\label{Fig:Noise}\textbf{Noise pollution analysis on the 60 km link.} Averaged correlation peaks across all waveplate settings in a full CHSH measurement, with (orange) and without (green) co-propagating classical light. The background noise, dominated by detector dark counts and thermal emission from the SRC, results in SNRs of 35 with classical ON, and 49 with classical OFF.}
\end{figure}

\newpage
Lastly, beyond classical-inspired KPIs, hybrid networks must also meet quantum-specific requirements, the most critical being low noise in the quantum measurements (see Sec.\ref{Sec:FWMSource}). In our setup, quantum signals are transmitted in the 1310~nm CWDM band, while classical traffic co-propagates in the C-band using a DWDM channel at 1560.61~nm (Channel 21 at +4~dBm launch power), combined via OADMs. Unlike the entangled photons, the background noise and stray classical light don't change significantly in intensity as the waveplate rotates before the polarizer, showing they’re mostly unpolarized. As shown in Figure \ref{Fig:Noise}, the 60~km path maintains similar noise levels when the quantum signal is distributed in the presence or absence of classical light, confirming the viability of co-propagation.

\section{Conclusion}
Practical quantum networking requires critical components including bright, deployable entanglement sources producing high-fidelity photon pairs, rapid reference frame stabilization, and low-loss routing. Here, we addressed these challenges and successfully deployed \textit{BearlinQ}, a quantum testbed in the metropolitan fiber network of Deutsche Telekom in Berlin.

Using commercially available devices and telecom-standard optical add-drop multiplexing hardware, we demonstrated uninterrupted high-rate entanglement distribution with quantum signals in the telecom 1324~nm O-band seamlessly coexisting with classical bidirectional C-band traffic along dynamically selectable fiber distances ranging from 10~m to 60~km. With active polarization stabilization, our system maintained high entanglement fidelity, with lower and upper bounds ranging from 85\% to 99\% and with $S$ values between 2.36 and 2.74, indicating strong quantum nonlocality, with near telecom-grade reliability ($< 1.5$\% downtime) over extended durations, all without requiring dedicated fiber or modifications to existing telecom infrastructure. Our deployment also validates the feasibility of quantum routing within existing telecom infrastructure, demonstrating telecom operational reliability requirements over typical metropolitan distances.

By leveraging the co-existence of usable quantum photons with standard classical telecom traffic, this work sets operational benchmarks necessary for commercially viable quantum entanglement-based services, eliminating the costs and complexity of new fiber installations. These results highlight that quantum networks can be integrated with existing assets and are ready to transition from lab-based demonstrations to reliable deployments. By embracing these innovations, operators can establish themselves as the backbone of quantum-enabled networks, laying the technical foundation for commercially viable quantum services for quantum communications, distributed quantum computing, and quantum sensing applications integrated into existing urban infrastructure.

\bibliographystyle{IEEEtran}
\bibliography{library}

\providecommand{\noopsort}[1]{}\providecommand{\singleletter}[1]{#1}%
\begin{thebibliography}{10}
\providecommand{\url}[1]{#1}
\csname url@samestyle\endcsname
\providecommand{\newblock}{\relax}
\providecommand{\bibinfo}[2]{#2}
\providecommand{\BIBentrySTDinterwordspacing}{\spaceskip=0pt\relax}
\providecommand{\BIBentryALTinterwordstretchfactor}{4}
\providecommand{\BIBentryALTinterwordspacing}{\spaceskip=\fontdimen2\font plus
\BIBentryALTinterwordstretchfactor\fontdimen3\font minus
  \fontdimen4\font\relax}
\providecommand{\BIBforeignlanguage}[2]{{%
\expandafter\ifx\csname l@#1\endcsname\relax
\typeout{** WARNING: IEEEtran.bst: No hyphenation pattern has been}%
\typeout{** loaded for the language `#1'. Using the pattern for}%
\typeout{** the default language instead.}%
\else
\language=\csname l@#1\endcsname
\fi
#2}}
\providecommand{\BIBdecl}{\relax}
\BIBdecl

\bibitem{Cacciapuoti2020}
A.~S. Cacciapuoti, M.~Caleffi, F.~Tafuri, F.~S. Cataliotti, S.~Gherardini, and
  G.~Bianchi, ``Quantum internet: Networking challenges in distributed quantum
  computing,'' \emph{IEEE Network}, vol.~34, no.~1, pp. 137--143, 2020.

\bibitem{Pogorelov2021}
\BIBentryALTinterwordspacing
I.~Pogorelov, T.~Feldker, C.~D. Marciniak, L.~Postler, G.~Jacob,
  O.~Krieglsteiner, V.~Podlesnic, M.~Meth, V.~Negnevitsky, M.~Stadler,
  B.~H\"ofer, C.~W\"achter, K.~Lakhmanskiy, R.~Blatt, P.~Schindler, and
  T.~Monz, ``Compact ion-trap quantum computing demonstrator,'' \emph{PRX
  Quantum}, vol.~2, p. 020343, Jun 2021. [Online]. Available:
  \url{https://link.aps.org/doi/10.1103/PRXQuantum.2.020343}
\BIBentrySTDinterwordspacing

\bibitem{QuEra}
\BIBentryALTinterwordspacing
D.~Bluvstein, H.~Levine, G.~Semeghini, T.~T. Wang, S.~Ebadi, M.~Kalinowski,
  A.~Keesling, N.~Maskara, H.~Pichler, M.~Greiner, V.~Vuleti{\'c}, and M.~D.
  Lukin, ``A quantum processor based on coherent transport of entangled atom
  arrays,'' \emph{Nature}, vol. 604, no. 7906, pp. 451--456, 2022. [Online].
  Available: \url{https://doi.org/10.1038/s41586-022-04592-6}
\BIBentrySTDinterwordspacing

\bibitem{Cacciapuoti2025}
A.~S. Cacciapuoti, C.~Pellitteri, J.~Illiano, L.~d'Avossa~amd Francesco~Mazza,
  S.~Chen, and M.~Caleffi, ``Quantum data centers: Why entanglement changes
  everything,'' \emph{arXiv}, 2025.

\bibitem{Komar2014}
\BIBentryALTinterwordspacing
P.~K{\'{o}}m{\'{a}}r, E.~M. Kessler, M.~Bishof, L.~Jiang, A.~S. S{\o}rensen,
  J.~Ye, and M.~D. Lukin, ``{A quantum network of clocks},'' \emph{Nature
  Physics}, vol.~10, no.~8, pp. 582--587, 2014. [Online]. Available:
  \url{https://doi.org/10.1038/nphys3000}
\BIBentrySTDinterwordspacing

\bibitem{Wu2019}
\BIBentryALTinterwordspacing
X.~Wu, Z.~Pagel, B.~S. Malek, T.~H. Nguyen, F.~Zi, D.~S. Scheirer, and
  H.~M{\"u}ller, ``Gravity surveys using a mobile atom interferometer,''
  \emph{Science Advances}, vol.~5, no.~9, p. eaax0800, 2019. [Online].
  Available: \url{https://www.science.org/doi/abs/10.1126/sciadv.aax0800}
\BIBentrySTDinterwordspacing

\bibitem{Kimble2008}
\BIBentryALTinterwordspacing
H.~J. Kimble, ``{The quantum internet},'' \emph{Nature}, vol. 453, no. 7198,
  pp. 1023--1030, 2008. [Online]. Available:
  \url{https://doi.org/10.1038/nature07127}
\BIBentrySTDinterwordspacing

\bibitem{Guo2020}
\BIBentryALTinterwordspacing
X.~Guo, C.~R. Breum, J.~Borregaard, S.~Izumi, M.~V. Larsen, T.~Gehring,
  M.~Christandl, J.~S. Neergaard-Nielsen, and U.~L. Andersen, ``Distributed
  quantum sensing in a continuous-variable entangled network,'' \emph{Nature
  Physics}, vol.~16, no.~3, pp. 281--284, 2020. [Online]. Available:
  \url{https://doi.org/10.1038/s41567-019-0743-x}
\BIBentrySTDinterwordspacing

\bibitem{Bhaskar2020}
\BIBentryALTinterwordspacing
M.~K. Bhaskar, R.~Riedinger, B.~Machielse, D.~S. Levonian, C.~T. Nguyen, E.~N.
  Knall, H.~Park, D.~Englund, M.~Lon{\v c}ar, D.~D. Sukachev, and M.~D. Lukin,
  ``Experimental demonstration of memory-enhanced quantum communication,''
  \emph{Nature}, vol. 580, no. 7801, pp. 60--64, 2020. [Online]. Available:
  \url{https://doi.org/10.1038/s41586-020-2103-5}
\BIBentrySTDinterwordspacing

\bibitem{Duan2001}
\BIBentryALTinterwordspacing
L.-M. Duan, M.~D. Lukin, J.~I. Cirac, and P.~Zoller, ``{Long-distance quantum
  communication with atomic ensembles and linear optics},'' \emph{Nature}, vol.
  414, no. 6862, pp. 413--418, 2001. [Online]. Available:
  \url{https://doi.org/10.1038/35106500}
\BIBentrySTDinterwordspacing

\bibitem{Jasminder2021}
\BIBentryALTinterwordspacing
J.~S. Sidhu, S.~K. Joshi, M.~G{\"u}ndoğan, T.~Brougham, D.~Lowndes,
  L.~Mazzarella, M.~Krutzik, S.~Mohapatra, D.~Dequal, G.~Vallone, P.~Villoresi,
  A.~Ling, T.~Jennewein, M.~Mohageg, J.~G. Rarity, I.~Fuentes, S.~Pirandola,
  and D.~K.~L. Oi, ``Advances in space quantum communications,'' \emph{IET
  Quantum Communication}, vol.~2, no.~4, pp. 182--217, 2021. [Online].
  Available:
  \url{https://ietresearch.onlinelibrary.wiley.com/doi/abs/10.1049/qtc2.12015}
\BIBentrySTDinterwordspacing

\bibitem{Knaut2024}
\BIBentryALTinterwordspacing
C.~M. Knaut, A.~Suleymanzade, Y.~C. Wei, D.~R. Assumpcao, P.~J. Stas, Y.~Q.
  Huan, B.~Machielse, E.~N. Knall, M.~Sutula, G.~Baranes, N.~Sinclair,
  C.~De-Eknamkul, D.~S. Levonian, M.~K. Bhaskar, H.~Park, M.~Lon{\v c}ar, and
  M.~D. Lukin, ``Entanglement of nanophotonic quantum memory nodes in a telecom
  network,'' \emph{Nature}, vol. 629, no. 8012, pp. 573--578, 2024. [Online].
  Available: \url{https://doi.org/10.1038/s41586-024-07252-z}
\BIBentrySTDinterwordspacing

\bibitem{Kucera2024}
S.~Kucera, .~C. Haen, .~E. Arensk{\"o}tter, .~T. Bauer, .~J. Meiers, .~M.
  Sch{\"a}fer, and .~Ross, ``Demonstration of quantum network protocols over a
  14-km urban fiber link,'' \emph{npj Qunatum Information}, vol.~10, p.~88,
  2024.

\bibitem{Tagliavacche2025}
N.~Tagliavacche, M.~Borghi, G.~Guarda, D.~Ribezzo, M.~Liscidini, D.~Bacco,
  M.~Galli, and D.~Bajoni, ``Frequency-bin entanglement-based quantum key
  distribution,'' \emph{npj Quantum Inf}, vol.~10, p.~60, 2025.

\bibitem{Aaronson2009a}
S.~Aaronson, ``Quantum copy-protection and quantum money,'' in \emph{24th
  Annual IEEE Conference on Computational Complexity}.\hskip 1em plus 0.5em
  minus 0.4em\relax IEEE, Jul. 2009, pp. 229--242.

\bibitem{Yuan2008}
Y.~LI and G.~Zeng, ``Quantum anonymous and voting systems and based on
  entangled state,'' \emph{Optical Review}, vol.~15, p. 219–223, 2008.

\bibitem{Briegel1998}
\BIBentryALTinterwordspacing
H.-J. Briegel, W.~D\"ur, J.~I. Cirac, and P.~Zoller, ``Quantum repeaters: The
  role of imperfect local operations in quantum communication,'' \emph{Phys.
  Rev. Lett.}, vol.~81, pp. 5932--5935, Dec 1998. [Online]. Available:
  \url{https://link.aps.org/doi/10.1103/PhysRevLett.81.5932}
\BIBentrySTDinterwordspacing

\bibitem{Graham2022}
\BIBentryALTinterwordspacing
T.~M. Graham, Y.~Song, J.~Scott, C.~Poole, L.~Phuttitarn, K.~Jooya, P.~Eichler,
  X.~Jiang, A.~Marra, B.~Grinkemeyer, M.~Kwon, M.~Ebert, J.~Cherek, M.~T.
  Lichtman, M.~Gillette, J.~Gilbert, D.~Bowman, T.~Ballance, C.~Campbell, E.~D.
  Dahl, O.~Crawford, N.~S. Blunt, B.~Rogers, T.~Noel, and M.~Saffman,
  ``{Multi-qubit entanglement and algorithms on a neutral-atom quantum
  computer},'' \emph{Nature}, vol. 604, no. 7906, pp. 457--462, 2022. [Online].
  Available: \url{https://doi.org/10.1038/s41586-022-04603-6}
\BIBentrySTDinterwordspacing

\bibitem{Quan2016}
R.~Quan, Y.~Zhai, M.~Wang, F.~Hou, S.~Wang, X.~Xiang, T.~Liu, S.~Zhang, and
  R.~Dong, ``Demonstration of quantum synchronization based on second-order
  quantum coherence of entangled photons,'' \emph{Scientific Reports}, vol.~6,
  no.~1, Jul. 2016.

\bibitem{strobel2024high}
\BIBentryALTinterwordspacing
T.~Strobel, S.~Kazmaier, T.~Bauer, M.~Sch\"{a}fer, A.~Choudhary, N.~L. Sharma,
  R.~Joos, C.~Nawrath, J.~H. Weber, W.~Nie, G.~Bhayani, L.~Wagner,
  A.~Bisquerra, M.~Geitz, R.-P. Braun, C.~Hopfmann, S.~L. Portalupi, C.~Becher,
  and P.~Michler, ``High-fidelity distribution of triggered
  polarization-entangled telecom photons via a 36 km intra-city fiber
  network,'' \emph{Optica Quantum}, vol.~2, no.~4, pp. 274--281, Aug 2024.
  [Online]. Available:
  \url{https://opg.optica.org/opticaq/abstract.cfm?URI=opticaq-2-4-274}
\BIBentrySTDinterwordspacing

\bibitem{Thomas2023designing}
\BIBentryALTinterwordspacing
J.~M. Thomas, G.~S. Kanter, and P.~Kumar, ``Designing noise-robust quantum
  networks coexisting in the classical fiber infrastructure,'' \emph{Opt.
  Express}, vol.~31, no.~26, pp. 43\,035--43\,047, Dec 2023. [Online].
  Available: \url{https://opg.optica.org/oe/abstract.cfm?URI=oe-31-26-43035}
\BIBentrySTDinterwordspacing

\bibitem{mao2018integrating}
\BIBentryALTinterwordspacing
Y.~Mao, B.-X. Wang, C.~Zhao, G.~Wang, R.~Wang, H.~Wang, F.~Zhou, J.~Nie,
  Q.~Chen, Y.~Zhao, Q.~Zhang, J.~Zhang, T.-Y. Chen, and J.-W. Pan,
  ``Integrating quantum key distribution with classical communications in
  backbone fiber network,'' \emph{Opt. Express}, vol.~26, no.~5, pp.
  6010--6020, Mar 2018. [Online]. Available:
  \url{https://opg.optica.org/oe/abstract.cfm?URI=oe-26-5-6010}
\BIBentrySTDinterwordspacing

\bibitem{thomas2024quantum}
J.~M. Thomas, F.~I. Yeh, J.~H. Chen, J.~J. Mambretti, S.~J. Kohlert, G.~S.
  Kanter, and P.~Kumar, ``Quantum teleportation coexisting with classical
  communications in optical fiber,'' \emph{Optica}, vol.~11, no.~12, pp.
  1700--1707, 2024.

\bibitem{marcikic2004distribution}
I.~Marcikic, H.~De~Riedmatten, W.~Tittel, H.~Zbinden, M.~Legr{\'e}, and
  N.~Gisin, ``Distribution of time-bin entangled qubits over 50 km of optical
  fiber,'' \emph{Physical review letters}, vol.~93, no.~18, p. 180502, 2004.

\bibitem{shen2022distributing}
\BIBentryALTinterwordspacing
L.~Shen, C.~H. Chow, J.~Y.~X. Peh, X.~J. Yeo, P.~K. Tan, and C.~Kurtsiefer,
  ``Distributing polarization-entangled photon pairs with high rate over long
  distances through standard telecommunication fiber,'' \emph{Phys. Rev.
  Appl.}, vol.~18, p. 044075, Oct 2022. [Online]. Available:
  \url{https://link.aps.org/doi/10.1103/PhysRevApplied.18.044075}
\BIBentrySTDinterwordspacing

\bibitem{rahmouni2024100}
\BIBentryALTinterwordspacing
A.~Rahmouni, P.~S. Kuo, Y.~S. Li-Baboud, I.~A. Burenkov, Y.~Shi, M.~V. Jabir,
  N.~Lal, D.~Reddy, M.~Merzouki, L.~Ma, A.~Battou, S.~V. Polyakov, O.~Slattery,
  and T.~Gerrits, ``100-km entanglement distribution with coexisting quantum
  and classical signals in a single fiber,'' \emph{J. Opt. Commun. Netw.},
  vol.~16, no.~8, pp. 781--787, Aug 2024. [Online]. Available:
  \url{https://opg.optica.org/jocn/abstract.cfm?URI=jocn-16-8-781}
\BIBentrySTDinterwordspacing

\bibitem{Yang2025}
\BIBentryALTinterwordspacing
R.~Yang, A.~Wonfor, V.~Jain, E.~Fazel, M.~Clark, R.~D. Oliveira, S.~Bahrani,
  A.~Mehrpooya, A.~Olianezhad, M.~Alhussein, S.~Joshi, J.~Rarity, R.~Penty,
  R.~Wang, and D.~Simeonidou, ``A uk nationwide heterogeneous quantum
  network,'' in \emph{Optical Fiber Communication Conference (OFC) Postdeadline
  Papers 2025}.\hskip 1em plus 0.5em minus 0.4em\relax Optica Publishing Group,
  2025, p. Th4C.7. [Online]. Available:
  \url{https://opg.optica.org/abstract.cfm?URI=OFC-2025-Th4C.7}
\BIBentrySTDinterwordspacing

\bibitem{Tchebotareva2019}
\BIBentryALTinterwordspacing
A.~Tchebotareva, S.~L.~N. Hermans, P.~C. Humphreys, D.~Voigt, P.~J. Harmsma,
  L.~K. Cheng, A.~L. Verlaan, N.~Dijkhuizen, W.~de~Jong, A.~Dr\'eau, and
  R.~Hanson, ``Entanglement between a diamond spin qubit and a photonic
  time-bin qubit at telecom wavelength,'' \emph{Phys. Rev. Lett.}, vol. 123, p.
  063601, Aug 2019. [Online]. Available:
  \url{https://link.aps.org/doi/10.1103/PhysRevLett.123.063601}
\BIBentrySTDinterwordspacing

\bibitem{source2024}
\BIBentryALTinterwordspacing
A.~N. Craddock, Y.~Wang, F.~Giraldo, R.~Sekelsky, M.~Flament, and M.~Namazi,
  ``High-rate subgigahertz-linewidth bichromatic entanglement source for
  quantum networking,'' \emph{Phys. Rev. Appl.}, vol.~21, p. 034012, Mar 2024.
  [Online]. Available:
  \url{https://link.aps.org/doi/10.1103/PhysRevApplied.21.034012}
\BIBentrySTDinterwordspacing

\bibitem{craddock2024automated}
A.~N. Craddock, A.~Lazenby, G.~B. Portmann, R.~Sekelsky, M.~Flament, and
  M.~Namazi, ``Automated distribution of polarization-entangled photons using
  deployed new york city fibers,'' \emph{PRX Quantum}, vol.~5, no.~3, p.
  030330, 2024.

\bibitem{Clauser1969}
\BIBentryALTinterwordspacing
J.~F. Clauser, M.~A. Horne, A.~Shimony, and R.~A. Holt, ``Proposed experiment
  to test local hidden-variable theories,'' \emph{Phys. Rev. Lett.}, vol.~23,
  pp. 880--884, Oct 1969. [Online]. Available:
  \url{https://link.aps.org/doi/10.1103/PhysRevLett.23.880}
\BIBentrySTDinterwordspacing

\bibitem{bloembergen1967stimulated}
N.~Bloembergen, ``The stimulated raman effect,'' \emph{American Journal of
  Physics}, vol.~35, no.~11, pp. 989--1023, 1967.

\bibitem{Davidson2021}
\BIBentryALTinterwordspacing
O.~Davidson, R.~Finkelstein, E.~Poem, and O.~Firstenberg, ``Bright multiplexed
  source of indistinguishable single photons with tunable ghz-bandwidth at room
  temperature,'' \emph{New Journal of Physics}, vol.~23, no.~7, p. 073050, jul
  2021. [Online]. Available: \url{https://dx.doi.org/10.1088/1367-2630/ac14ab}
\BIBentrySTDinterwordspacing

\bibitem{Ren1988}
\BIBentryALTinterwordspacing
Z.~B. Ren, P.~Robert, and P.-A. Paratte, ``Temperature dependence of bend- and
  twist-induced birefringence in a low-birefringence fiber,'' \emph{Opt. Lett.
  13, 62-64}, 1988. [Online]. Available:
  \url{https://opg.optica.org/ol/abstract.cfm?URI=ol-13-1-62}
\BIBentrySTDinterwordspacing

\bibitem{Day1983}
\BIBentryALTinterwordspacing
\emph{Birefringence measurements in single mode optical fiber}, vol.
  0425.\hskip 1em plus 0.5em minus 0.4em\relax SPIE, 1983. [Online]. Available:
  \url{https://doi.org/10.1117/12.936216}
\BIBentrySTDinterwordspacing

\bibitem{Sena2024}
M.~Sena, M.~Flament, M.~Namazi, S.~Andrewski, G.~Portmann, R.-P. Braun,
  M.~Youssef-Sayed, R.~D{\"o}ring, M.~Ritter, O.~Holschke \emph{et~al.},
  ``High-fidelity entanglement distribution through berlin using an
  operator’s fiber infrastructure,'' in \emph{Optical Fiber Communication
  Conference}.\hskip 1em plus 0.5em minus 0.4em\relax Optica Publishing Group,
  2025, pp. M4E--2.

\bibitem{semenov2010entanglement}
\BIBentryALTinterwordspacing
A.~A. Semenov and W.~Vogel, ``Entanglement transfer through the turbulent
  atmosphere,'' \emph{Phys. Rev. A}, vol.~81, p. 023835, Feb 2010. [Online].
  Available: \url{https://link.aps.org/doi/10.1103/PhysRevA.81.023835}
\BIBentrySTDinterwordspacing

\bibitem{James2001}
\BIBentryALTinterwordspacing
D.~F.~V. James, P.~G. Kwiat, W.~J. Munro, and A.~G. White, ``Measurement of
  qubits,'' \emph{Phys. Rev. A}, vol.~64, p. 052312, Oct 2001. [Online].
  Available: \url{https://link.aps.org/doi/10.1103/PhysRevA.64.052312}
\BIBentrySTDinterwordspacing

\end{thebibliography}

\newpage
\section*{Supplementary Material}
\subsection{Performance Optimizations}
\label{Sec:SupMattPerf}
\begin{figure}[h]
\centering
\includegraphics[width=\columnwidth]{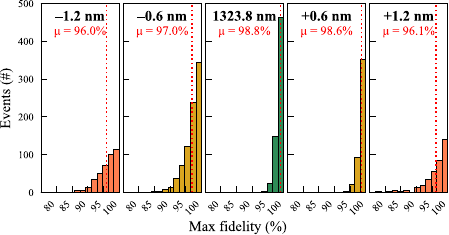}
\caption{\label{Fig:fidelityloss}\textbf{APC wavelength impact on quantum fidelity}. As probing wavelength deviates from that of qubits, the classically inferred fidelity becomes increasingly unreliable. The discrepancy grows with spectral detuning, underscoring the importance of a near-zero offset to ensure accurate $F_Q$. This is further exacerbated by PMD effects in long fibers.}
\end{figure}

\textbf{Fidelity improvements}: High-fidelity entanglement distribution relies on spectrally narrowband photons to minimize internal wavelength spread and associated polarization dispersion. However, quantum fidelity is fundamentally constrained when the APC is calibrated at a wavelength different from that of the quantum signal because fiber-induced polarization rotations are both wavelength-dependent and time-varying. Measurements in deployed telecom fibers show that the fiber's polarization transformation varies significantly with wavelength. This wavelength dependence becomes pronounced over tens of kilometers, rendering a system calibrated at one wavelength inherently misaligned for others. When polarization compensation is performed at 1324~nm, high fidelity is maintained only for photons within a narrow spectral range (<0.5~nm). As shown in Figure~\ref{Fig:fidelityloss}, even slight detuning (or a broad photon bandwidth) introduces measurable infidelities. Our data show that spectral mismatch consistently reduces Bell-state fidelity across all fiber configurations. This reduction was due to high fiber attenuation that limited the operation of APC at the target wavelength due to insufficient optical power, leading to quantum fidelity losses of up to 5\%. This constraint can be mitigated.

\subsection{Fidelity Calculations}
\label{Sec:AppendixFidelity}
Despite not performing full quantum state tomography \cite{James2001}, we can still estimate approximate quantum fidelity bounds based on the assumption that the generated state is close to a maximally entangled
 Bell-state: $\ket{\phi_+}=\frac{1}{\sqrt{2}}(\ket{HH}+\ket{VV})$. The fidelity between the reconstructed density matrix and the target state is given by
{\begin{align}
 F &= Tr[\ket{\phi_+}\bra{\phi_+}\rho\ket{\phi_+}\bra{\phi_+}] \\
 &= \: \frac{\rho_{11}+\rho_{44}+\rho_{41}+\rho_{14}}{2}\label{eq:f}
\end{align}}

\noindent where $\rho$ is the density matrix of the two-photon state in the $\{\ket{HH}, \ket{HV}, \ket{VH}, \ket{VV}\}$ basis. The diagonal terms $\rho_{11}$ and $\rho_{44}$ correspond to the probabilities of measuring $\ket{HH}$, $\ket{VV}$ states and the off-diagonal terms $\rho_{14}$, $\rho_{41}=\rho^*_{14}$ represent the coherence between these components. We define the normalization constant, $N = C_{HH} + C_{VV} + C_{DD} + C_{AA}$ where $C_{ij}$ is the number of coincidences in the $\ket{ij}$ basis. Diagonal elements are proportional to the respective coincidence counts
{\begin{equation}
 \rho_{11} = \frac{C_{HH}}{N},\; \rho_{44} = \frac{C_{VV}}{N}
\end{equation}}

\noindent Coincidence measurements in the diagonal and anti-diagonal basis relate to the off-diagonal elements by
{\begin{equation}
 \frac{C_{DD}+C_{AA}}{N} = \frac{1 + \rho_{14}+ \rho_{23}+ \rho_{32}+ \rho_{41}}{2}
\end{equation}}

\noindent Substituting this into \ref{eq:f} gives
{\begin{equation}
 F = \frac{C_{HH} + C_{VV} + 2C_{DD} + 2C_{AA}}{2N} - \frac{1 + \rho_{23} + \rho_{32}}{2}
\end{equation}}

\noindent We can bound the final term by noting that $|\rho_{23}|= |\rho_{32}| \leq \sqrt{\rho_{22}\rho_{33}}$ such that we obtain
{\begin{equation}
\begin{array}{c}
\displaystyle
\frac{C_{HH} + C_{VV} + 2C_{DD} + 2C_{AA} - 2\sqrt{C_{HV} C_{VH}}}{2N} - \frac{1}{2} \\[1.0em]
\leq F \leq \\[0.5em]
\displaystyle
\frac{C_{HH} + C_{VV} + 2C_{DD} + 2C_{AA} + 2\sqrt{C_{HV} C_{VH}}}{2N} - \frac{1}{2}
\end{array}
\end{equation}}

\noindent Similarly, we obtain the second inequality
{\begin{equation}
\begin{array}{c}
\displaystyle
\frac{C_{DD} + C_{AA} + 2C_{HH} + 2C_{VV} - 2\sqrt{C_{AD} C_{DA}}}{2N} - \frac{1}{2} \\[1.0em]
\leq F \leq \\[0.5em]
\displaystyle
\frac{C_{DD} + C_{AA} + 2C_{HH} + 2C_{VV} + 2\sqrt{C_{AD} C_{DA}}}{2N} - \frac{1}{2}
\end{array}
\end{equation}}

\noindent The lower minimum (upper maximum) bounds of the two inequalities are taken as the lower (upper) bounds of the entanglement fidelity.

\end{document}